\documentclass[12pt]{article}
\usepackage{latexsym,amsmath,amsfonts,amssymb}
\usepackage[latin1]{inputenc}
\usepackage[pdftex]{graphicx}

\usepackage{euscript,mathrsfs}
\usepackage{fullpage}
\usepackage{color}
\usepackage{bbm}
\usepackage[american]{babel}

\renewcommand{\baselinestretch}{1.2}
\newcommand{\be}{\begin{equation}}
\newcommand{\ee}{\end{equation}}
\newcommand{\bea}{\begin{equation} \begin{aligned}} 
\newcommand{\eea}{\end{aligned} \end{equation}}
\newcommand{\rref}[1]{(\ref{#1})}

\newcommand{\comment}[1]{}

\newcommand{\ads}{\mathnormal{AdS}}

\newcommand{\p}{\partial}

\newcommand{\bX}{{\bf X}}

\newcommand{\bE}{{\bf E}}
\newcommand{\bY}{{\bf Y}}

\newcommand{\m}{\mu}

\newcommand{\s}{\sigma}
\renewcommand{\t}{\tau}
\renewcommand{\a}{\alpha}

\renewcommand{\l}{\lambda}

\renewcommand{\ads}{AdS }

\newcommand{\R}{\mathbb{R}}

\renewcommand{\H}{\mathbb{H}}

\begin{document}

\begin{titlepage}

\

\

\vfil
\begin{center}
{\LARGE Topologically Massive Gravity and Ricci-Cotton Flow} 
\end{center}

\vfil
\begin{center}
{Nima Lashkari \& Alexander Maloney}\let\thefootnote\relax\footnotetext{emails: {\tt lashkari@physics.mcgill.ca, maloney@physics.mcgill.ca}}\vspace{6.0mm}\\

\

McGill Physics Department, 3600 rue University, Montreal, QC H3A 2T8, Canada\\
\end{center}
\vspace{3mm}

\vfil
\begin{center}
{\large Abstract}
\end{center}
\noindent
We consider Topologically Massive Gravity (TMG), which is three dimensional general relativity with a cosmological constant and a gravitational Chern-Simons term.  When the cosmological constant is negative the theory has two potential vacuum solutions: Anti-de Sitter space and Warped Anti-de Sitter space.  The theory also contains a massive graviton state which renders these solutions unstable for certain values of the parameters and boundary conditions.  We study the decay of these solutions due to the condensation of the massive graviton mode using Ricci-Cotton flow, which is the appropriate generalization of Ricci flow to TMG.  When the Chern-Simons coupling is small the AdS solution flows to warped AdS by the condensation of the massive graviton mode.  When the coupling is large the situation is reversed, and warped AdS flows to AdS.   Minisuperspace models are constructed where these flows are studied explicitly. 

\vfil
\end{titlepage}
\newpage

\renewcommand{\baselinestretch}{1.1}  
\renewcommand{\arraystretch}{1.5}

\section{Introduction}

Three dimensional gravity has proven a fruitful and fascinating subject where basic conceptual questions in quantum gravity can be addressed in a simple  -- and possibly exactly solvable -- toy model.  Recently attention has focused on a particular class of theories of three dimensional gravity: Einstein gravity with a negative cosmological constant and a gravitational Chern-Simons term.  This theory is known as Topologically Massive Gravity (TMG) \cite{Deser:1981wh,Deser:1982vy}.  TMG has a rich space of solutions, including both black holes and perturbative gravitons, but nevertheless has the potential to be exactly solvable\footnote{For a discussion of solution generating techniques in TMG see \cite{Nutku:1993eb,Gurses:2010sm}.}. Indeed, for a special ``chiral'' value of the Chern-Simons coupling constant -- and with certain boundary conditions -- there is evidence that the theory can be quantized and the exact spectrum of the theory computed \cite{Li:2008dq, Maloney:2009ck}.  

For more general values of the coupling constants, however, the situation is less clear.  This is because, unlike the case of three dimensional Einstein gravity, TMG is not locally trivial.   In addition to locally Anti-de Sitter (AdS) solutions, the theory possesses solutions which are locally warped AdS.  These warped AdS geometries are fibres of circles or lines over AdS${}_2$ or $\H_2$ and come in four varieties: they are space like or time-like depending on the signature of the fibre, and squashed or stretched depending on the value of Chern-Simons coupling.  These warped geometries share many features with unwarped AdS and provide a possible new setting for the holographic correspondence that goes beyond standard AdS/CFT.  Higher dimensional versions of these space-times have recently been studied as potential duals to non-relativistic systems of interest in condensed matter physics \cite{Balasubramanian:2008dm, Son:2008ye}.  Moreover, warped AdS arises as a fibre in the near-horizon geometry of the four dimensional extremal Kerr black hole and appears to play a crucial role in the conjectured Kerr/CFT correspondence \cite{Guica:2008mu}.  Clearly an understanding of holography for these geometries would represent significant progress.  Indeed, based on the black hole entropy formula it was conjectured that the spacelike stretched version of warped AdS is dual to a CFT \cite{Anninos:2008fx}.

In attempting to formulate a holographic correspondence for these backgrounds, one immediately confronts the problem that for certain values of the coupling constant the AdS (and possibly the warped AdS) solutions described above are unstable. In particular, the local graviton excitations around AdS have negative energy (see e.g. \cite{Li:2008dq}).  This will cause the AdS solution to decay.  The goal of this paper is to study this decay process.  The AdS and warped AdS solutions described above are the most symmetric solutions of the equations of motion so it is natural to guess that they are connected by the condensation of this massive graviton mode.  We will argue that this is indeed the case.

Of course, it is difficult to study explicitly the non-linear evolution of metric modes describing this decay process.  We will therefore use an alternative approach, based on string worldsheet techniques.  There is considerable evidence that renormalization group flow on the string worldsheet is equivalent to the condensation of a space-time tachyon (see e.g. {\cite{Headrick:2004hz} and references therein).  Under such a flow the space-time metric evolves according to the first order equation
\be\label{beta}
\frac{d g_{\mu\nu}}{dt} = \beta_{\mu\nu}
\ee
where $\beta_{\mu\nu}$ is the metric equation of motion and $t$ a renormalization scale parameter.  Fixed points of this flow are solutions of the space-time equations of motion. If an unstable solution is perturbed slightly, it will flow either to another solution or to a singular configuration of some sort.  The intermediate points in this flow may or may not be related to the detailed structure of the corresponding tachyon condensation process, but the starting and end points of this renormalization flow should coincide with those of the corresponding tachyon condensation process.  This provides a significant technical advantage, as the renormalization group equation is first order in the metric while field equations are (at least)  second order. 

In standard Einstein-gravity, the right hand side of equation \rref{beta} is a linear combination of the Ricci tensor and the metric, and the resulting equations describe Ricci flow.  This can be seen explicitly by computing the worldsheet beta function for the space-time metric. In the present case, however,  we will not focus on a specific string worldsheet theory. \footnote{Of course one could try to be more ambitious and study the tachyon condensation using worldsheet techniques.  String compactifications to \ads with gravitational Chern-Simons terms are easy to construct.  However, they typically contain Ramond-Ramond fields which preclude a simple worldsheet construction.} 
 Instead we will simply replace the right hand side by the appropriate metric equation of motion.  The right hand side of equation \rref{beta} is then a linear combination of the metric, Ricci and Cotton tensors and the resulting flow is known as Ricci-Cotton flow\footnote{For a discussion of Cotton flow see \cite{Kisisel:2008jx}.}.  
 
In this paper we will use Ricci-Cotton flow to study transitions  between the AdS and warped AdS vacuum geometries.  Of course, the flow \rref{beta} is quite complicated, so we will start by considering the restriction of the flow to families of metrics with two commuting Killing vectors.  This family of metrics includes both AdS and warped AdS, so it is natural to look for flows between these two geometries that live within this class of metrics.  Indeed, there is a rich family of solutions to TMG within this class (see e.g. \cite{Ertl:2010dh} for a recent discussion).  We will see that for this class of metrics there exist explicit flows that connect warped AdS with AdS.  Thus AdS and warped AdS are connected by the condensation of the massive graviton mode.  

Before describing the details, let us briefly summarize our results.  The action of Topologically Massive Gravity
 \be\label{actionTMG}
S = \frac{1}{16\pi G}\left[\int d^3x\sqrt{-g}(R+2/\ell^2)+{1
\over \mu}S_{CS}\right] \ee 
 has three coupling constants: Newton's constant $G$, a cosmological constant $-2/\ell^2$ and a parameter $\mu$ which determines the size of the gravitational Chern-Simons term.   The curvature radius of the AdS solution is $\ell$.  We will see that when the coefficient of the Chern-Simons term is smaller than a critical value ($\mu \ell > 3$) the AdS vacuum flows to warped (stretched) AdS.    Thus, when perturbed in an appropriate manner, AdS will decay to warped AdS.  This is consistent with the conjecture of \cite{Anninos:2008fx} that spacelike stretched AdS is dual to a CFT when {$\mu \ell > 3$}.  Moreover, it indicates that AdS should arise as an unstable saddle point  -- a "sphaleron" configuration -- of the warped AdS theory.   


For smaller values ($\mu \ell <3$) the situation is reversed.  The warped solutions are squashed rather than stretched, and flow towards AdS under the appropriate Ricci-Cotton flow.  As there is no conjecture for a CFT dual to squashed AdS, the physical implications of this result are less clear.

In section 2 we describe the Ricci-Cotton flow between solutions of TMG. In section 3 we restrict our attention to metrics with two Killing vectors and show that the flow takes the form of a heat equation on $\R^{2,1}$.  In section 4 we construct a minisuperspace model in which AdS and warped AdS are connected by Ricci-Cotton flow.  

\section{Ricci-Cotton Flow and TMG}

In this section we describe a few basic features of TMG and its relation to Ricci-Cotton flow.

\subsection{Topologically Massive Gravity}

The action of Topologically Massive Gravity is\footnote{In this paper we use the conventional positive sign for Newton's constant.  With this sign black holes have positive energy.} \be\label{action}
S = \frac{1}{16\pi G}\left[\int d^3x\sqrt{-g}(R+2/\ell^2)+{1
\over \mu}S_{CS}\right] \ee where $S_{CS}$ is the gravitational
Chern-Simons term \be\label{csaction} S_{CS} = \frac{1}{2 }\int_{\mathcal{M}}
d^3x\sqrt{-g}
\varepsilon^ {\lambda\mu\nu}\Gamma^{r}_{\lambda
\sigma}\left(\partial_{\mu}\Gamma^ \sigma_{{r}
\nu}+\frac{2}{3}\Gamma^\sigma_{\mu\tau}\Gamma^\tau_{\nu {r}} \right)
\ee
We will use units with $\ell=1$.
The variation of the action with respect to the metric is \be\label{eom}\frac{\delta S}{\delta g^{\mu\nu}}=
{1\over 16 \pi G} E_{\mu\nu},~~~~~E_{\mu\nu}\equiv G_{\mu\nu}-{1\over \ell^2} g_{\mu\nu}+{1\over\mu}C_{\mu\nu} \ee The Cotton tensor \be\label{cgdef} C_{\mu\nu}\equiv \epsilon^{\alpha\beta}\,_{\mu}
\nabla_\a\left(R_{\nu\beta}-{1\over 4} R g_{\nu\beta}\right)\ee arises from the variation of $S_{CS}$.  The Cotton tensor is parity-odd, since the Chern-Simons Lagrangian is a pseudo-scalar rather than a scalar.  Thus a parity reversing diffeomorphism is equivalent to  taking $\mu \to -\mu$.  We will take $\mu>0$. The Cotton tensor is traceless by the Bianchi identity, so the scalar curvature of solutions is fixed by the trace of equations of motion to be $R=-6$.

The equations of motion $E_{\mu\nu}=0$ are solved by AdS${}_3$, which has metric
\be\label{ads}
ds^2=-\cosh(\rho)^2d\tau^2+\sinh(\rho)^2d\phi^2+d\rho^2
 \ee
This is the maximally symmetric Einstein space, with $SL(2,\R)\times SL(2,\R)$ isometry group and vanishing Cotton tensor.  The spectrum of gravitational perturbations of TMG in AdS has been studied extensively.  In addition to the usual massless gravitons, the theory contains a massive graviton state from which TMG gets its name. These massive graviton excitations can be organized into $SL(2,\R)_L\times SL(2,\R)_R$ representation with weights \cite{Li:2008dq} \footnote{There exists another massive mode with weight $\left(\frac{(3-\mu)}{2},-\frac{(1+\mu)}{2}\right)$ which is discarded to its divergence at the boundary.}
\be\label{gravitons}
(h,{\bar h})=\left(\frac{3+\mu}{2},\frac{\mu-1}{2}\right)
\ee 
%
The energy of this mode was computed in  \cite{Li:2008dq} and shown to be negative for values of $\mu\ne1$. Thus AdS is expected to be unstable away from this critical value.  At $\mu=1$ the mode structure is more complicated and whether or not the theory is stable depends in detail on the choice of boundary conditions; in this paper we focus on the $\mu\ne 1$ case.


The theory in addition possesses solutions with non-vanishing Cotton tensor.  The prototypical such solution is warped AdS:
\be\label{wads}
 ds^2= \frac{1}{\nu^2+3}\left(\pm(1+\s^2)(-d\tau^2)+\frac{d\s^2}{1+\s^2}\pm\frac{4\nu^2}{\nu^2+3}(d\phi+\s d\tau)^2\right)\ee
where $\nu=\mu/3$. Warped AdS comes in two varieties: the positive (negative) sign refers to spacelike (timelike) warped AdS.\footnote{In addition one can construct a null warped AdS, which is a solution at the critical value of $\mu=3$.  Although we do not focus on null warped AdS here it might have an interesting role to play.  Tachyon condensation along a null direction can often be described exactly,  as in \cite{Hellerman:2006nx}. }  For spacelike (timelike) warped AdS the $(\tau,\sigma)$ directions describe an AdS${}_2$ ($\H_2$), and the $\phi$ direction describes a spacelike (timelike) line bundle fibered over this AdS${}_2$ ($\H_2$) base space.  When $\mu>3$ these solutions are referred to as stretched and for $\mu<3$ they are squashed.
 The metric \rref{wads} has $SL(2,\R)\times U(1)$ isometry group. 
 
 The stability of warped AdS has been less well studied in the literature.  Timelike warped AdS contains closed timelike curves and is expected to be unstable.  The stability of spacelike AdS, however, depends on the choice of boundary conditions.  For certain boundary conditions the spectrum was computed in \cite{Anninos:2009zi}, and it was argued that spacelike AdS is stable when $\mu>3$.  
This is consistent with the conjecture of \cite{Anninos:2008fx} that for $\mu>3$ spacelike warped AdS possesses a CFT dual (see also \cite{Compere:2007in,Compere:2008cv,Compere:2009zj}).

\subsection{Ricci-Cotton Flow}

Our goal is to understand the condensation of the massive graviton mode in TMG. To do so we will use Ricci-Cotton flow, which is a generalization of Ricci flow.  We refer the reader to  \cite{Headrick:2006ti}  and references therein for more details on Ricci flow and its relation to tachyon condensation.

Ricci-Cotton flow is a first order flow in the space of metrics.  To describe this flow, we start by considering the action of TMG as a function on the space of metrics.  Ricci-Cotton flow is the gradient flow with respect to this action.  Solutions to the equations of motion will, by definition, be fixed points of this flow.  More explicitly, we consider a family of metrics $g_{\mu\nu}(t)$ which depend smoothly on a parameter $t$.  This parameter $t$ labels different metrics and should not be confused with one of the space-time coordinates.   The Ricci-Cotton flow equation is 
\begin{equation}\label{flow}
 \frac{dg^A(t)}{dt}=-G^{AB}\frac{\p S}{\p g^B},
\end{equation} 
where $g^A$ denotes a metric and $S$ the action.  Here $A$ is an abstract index which includes both the covariant indices $\mu\nu$ and space-time point $x^\mu$ of the metric $g_{\mu\nu}$.  In order to define this flow we are required to specify a metric $G_{AB}$ on the space of metrics.
The most general local, diffeomorphism invariant such metric is 
\begin{equation}\label{metric}
 G_{AB}dg^Adg^B=\frac{1}{32\pi G_N}\int_M\sqrt{g}\left(dg^\mu_{\:\nu}dg^\nu_{\:\mu}+a (dg^\mu_{\:\mu})\right),
\end{equation} 
where $dg^A$ is a one-form on the space of metrics.  This metric depends on a choice of a parameter $a$ which is, at this point, undetermined.

It is now worth pausing to discuss the string worldsheet interpretation of the gradient flow (\ref{flow}).  For a string propagating in a given space-time background, one may regard the space-time metric $g_{\mu\nu}$  as a collection of coupling constants in the 1+1 dimensional string worldsheet theory.    The beta function for the space-time metric $g_{\mu\nu}$ is, famously, equal to the space-time equation of motion for the metric.  So the flow (\ref{flow}) describes renormalization group flow on the string worldsheet.    A fixed point of this flow is a conformal worldsheet theory, which describes a solution to the space-time equations of motion.  This is most commonly discussed for strings propagating in flat space, where the one loop beta functions include the linearized Ricci and metric tensors as well as the stress tensors of the various matter fields present.   Higher loop terms may be computed as well, and lead to higher derivative terms in the effective action of string theory expanded around a flat background.  Although typically discussed in the context of flat space, the gradient flow (\ref{flow}) is equivalent to renormalization group flow for more general space-times as well, provided there is a good worldsheet description.  
The RG flow interpretation also has the advantage that it fixes the undetermined parameter $a$ in the metric (\ref{metric}) on the space of metrics.  The one loop beta function of string theory gives  $a=-1/2$ \cite{Headrick:2006ti}; we will use this value from now on.

In this string theory interpretation,  the parameter $t$ is related to the renormalization scale of the worldsheet theory.  We typically normalize so that  $t=-1/2 \alpha'\log \Lambda$ where $\Lambda$ is the energy scale of the worldsheet theory.  Thus RG flow towards the infrared of the worldsheet theory is flow forward in time $t$.  In fact, this RG flow can be used to describe tachyon condensation.  To see this, consider a solution to the space-time equations of motion which possesses a tachyon.  In the worldsheet interpretation, this means that we have a  conformal field theory which  possesses a relevant operator.  We will perturb the solution by turning on this relevant tachyon operator on the worldsheet.  This breaks conformal invariance and causes the theory to flow to a new conformal theory in the IR.  So RG flow takes us from one solution to the space-time equations of motion in the UV to another solution in the IR.  In fact, the endpoint of this RG flow is argued to be precisely the endpoint of the corresponding space-time tachyon condensation.
We  refer the reader to the literature  for further details on the relationship between RG flow and tachyon condensation (e.g. \cite{Adams:2001sv, Gutperle:2002ki, Freedman:2005wx}).   

We can now write down the flow which arises in topologically massive gravity by setting the action in (\ref{flow}) to be that of TMG.
From the equation of motion \rref{eom} we find the Ricci-Cotton  flow equation
\begin{eqnarray}
 {\dot g}_{\mu\nu}&=&-2E_{\mu\nu}+2g_{\mu\nu}E^\alpha{}_{\alpha}\nonumber\\
&=&-2(R_{\mu\nu}+2g_{\mu\nu}+\frac{1}{\mu}C_{\mu\nu}).
\end{eqnarray} 
Here $\cdot$ denotes $d/dt$.  This is a version of the Ricci-Hamilton-deTurck flow \cite{ricci}.
All of the solutions of TMG described in the previous section are fixed points of this flow.  In fact, equation (\ref{run}) may be generalized to
\be\label{run}
\dot g_{\mu\nu} = -2(R_{\mu\nu}+2g_{\mu\nu}+\frac{1}{\mu}C_{\mu\nu}) + \nabla_{(\mu}v_{\nu)}
\ee
for an arbitrary vector $v^\mu(t)$.  This additional term does not change the geometry of the resulting flow; it simply applies to the metric a $t$-dependent diffeomorphism generated by the vector $v^\mu$.

\section{The $U(1)\times U(1)$ Sector}

The Ricci-Cotton flow (\ref{run}) which describes decay processes in TMG is quite complicated. In this section we will therefore focus on a particular class of geometries -- those with two commuting Killing vectors -- where these equations simplify.  With an appropriate gauge choice we will see that the Ricci-Cotton equations reduce to a modified heat equation for a vector field in $\R^{2,1}$.  In the next section we will use these results to describe flows relating AdS and warped AdS.

\subsection{Ansatz and Reduced Action}
A three dimensional spacetime with two commuting $U(1)$ isometries may, by a choice of coordinates, be written as
 \begin{equation} \label{ansatz}{ds^2}=-X^+(\sigma)d\tau^2+X^-(\sigma)d\phi^2+2Y(\sigma)d\t d\phi+{d\sigma^2\over
 X^+(\sigma) X^-(\sigma)+Y(\sigma)^2}\end{equation}
The two $U(1)$ isometries are generated by Killing vectors $\p_\tau$ and $\p_\phi$.
The coefficient of $d\sigma^2$ is just a coordinate choice; it has been chosen so that $\sqrt{-g}=1$.   The utility of this parameterization is that the equations of motion reduce to that of a particularly simple one dimensional system. This parameterization has been previously used in a similar context in \cite{Nutku:1993eb, Clement:1994sb, Moussa:2008sj}.

The three functions $X^\pm(\sigma), Y(\sigma)$ can be packaged into a three dimensional vector $ {\bf X}$ with components $X^i$, $i=0,1,2$ given by
\begin{equation}
X^0 = {X^+(\sigma) - X^-(\sigma)\over 2},~~~~~
X^1 = {X^+(\sigma) + X^-(\sigma)\over 2},~~~~~X^2=Y
\end{equation}
The dynamics of this class of metrics can then be thought of as that of a particle with position ${\bf X}(\sigma)$ moving in the auxiliary space $\R^{2,1}$ parameterized by $\bX$.

In fact, the equations of motion are invariant under Lorentz transformation in this auxiliary $\R^{2,1}$.  To see this, note that in writing (\ref{ansatz}) we have not completely exhausted our diffeomorphism invariance.  For any $M \in SL(2,\R)$ we may consider the change of coordinates
\begin{equation}
\left({t\atop \phi}\right) \to M \left({t\atop \phi}\right),~~~~~~
\ee
which takes
\be
\left({X^+~~Y\atop Y~~X^-}\right)\to
M^T\left({X^+~~Y\atop Y~~X^-}\right)M ,
\end{equation}
This preserves the form of the metric (\ref{ansatz}).  
Using the usual map of $SL(2,\R)$ onto $SO(2,1)$ we see that
the metric (\ref{ansatz}) is preserved by those residual diffeomorphisms which take
\begin{equation}\label{lorentz}
{\bf X} \to \Lambda \cdot {\bf X}
\end{equation}
where $\Lambda\in SO(2,1)$ preserves the norm
\begin{equation}
{\bf X}^2 = \eta_{ij} X^i X^j = X^+ X^- +Y^2
\end{equation}
Here $\eta_{ij}$ is the flat Minkowski metric on $\R^{2,1}$.
These residual diffeomorphisms are the Lorentz transformations of the auxiliary three dimensional space $\R^{2,1}$ parameterized by $\bf X$. 
We will find it useful to define the Lorentz invariant dot product and cross product 
\be \textbf{A}\cdot
\textbf{B}\equiv\eta_{ij}A^iB^j,~~~~\hbox{and}~~~~~
(\textbf{A}\times \textbf{B})^i\equiv
\eta^{il}\epsilon_{ljk}A^jB^k,\quad \epsilon_{012}=1\ee

Inserting our ansatz (\ref{ansatz}) into the action of TMG we find 
\begin{equation} S={1\over16\pi G}\int d\sigma
\left({1\over 2}\bX'^2-{1\over2\mu}\bX\cdot(\bX'\times
\bX'')+2\right)\label{action2}\end{equation}
where $'$ denotes $\p_\sigma$.
This is the action of a Lorentz invariant particle mechanics in $\R^{2,1}$.
The variation of the action with respect to $\bX$ is 
\be
\frac{\delta S}{\delta \bX} = {1\over 16 \pi G} \bE,~~~~~\bE = \bX''+\frac{1}{2\mu} \left(2 \bX \times \bX''' + 3 \bX'\times \bX''\right)
\ee
The action described above is invariant under translations in $\sigma$, so one may construct the corresponding conserved Hamiltonian
\be
H = {1\over 2} \textbf{X}'^2 -{1\over \mu} \textbf{X}\cdot(\textbf{X}'\times\textbf{X}'')-2
\ee
One can also construct a conserved angular momentum associated with $SO(2,1)$ rotations, although we shall not do so here. 

For the ansatz (\ref{ansatz}), the equations of motion of TMG are
\be
\bE=0,~~~~~H=0
\ee
The second equation arises because in writing the action (\ref{action2}) we have gauge fixed the metric, so we must impose an additional equation of motion coming from the gauge-fixed degrees of freedom. 



\subsection{Ricci-Cotton Flow}

We will now write down the Ricci-Cotton flow (\ref{run}) explicitly for the class of solutions with $U(1)\times U(1)$ symmetry.

We proceed by plugging the ansatz \rref{ansatz} into the flow equation \rref{run}.  In doing so, however, it is important to note that we have made a coordinate choice in writing down the metric \rref{ansatz}.  The addition of the term $\nabla_{(\mu}v_{\nu)}$ in (\ref{run}) -- representing a $t$-dependent diffeomorphism applied to the metric -- is therefore crucial.  The vector $v^\mu$ must be chosen to ensure that the gauge choice (\ref{ansatz}) is preserved under the flow.  Among other things, this means that the $v^\mu$ is chosen so that the volume element $\sqrt{g}$ remains constant under the flow; a similar strategy was adopted in the case of Ricci flow by \cite{isenberg}.

We start by demanding that the flow does  not generate $d\t d\s $ or $d\phi d\s$ terms in the metric, i.e. that
\be
\dot g_{\t\sigma} = 0,~~~~~\dot g_{\phi \sigma} = 0.
\ee 
This implies that $v^\mu$ takes the form
\be
v^\mu\p_\mu = (Y^2 \t + (-Y^0+Y^1)\phi) \p_\t + ((Y^0+Y^1)\t - Y^2 \phi)\p_\phi + f(\sigma) \p_\sigma.
\ee
Here $Y^0$, $Y^1$ and $Y^2$ are independent of $\sigma$, but may be arbitrary functions of $t$.  It is useful to package them into the vector
\be
\bY = (Y^0, Y^1, Y^2)
\ee
Likewise the function $f(\sigma)$ can depend on  $t$ as well.  

We must also demand that $g_{\sigma\sigma}=1/\bX^2$ according to (\ref{ansatz}), which is equivalent to demanding that $\sqrt{-g}$ is constant as we evolve in $t$.  Imposing
\be
g^{\mu\nu}\dot g_{\mu\nu}=0,
\ee
we find that
\bea\label{feq}
f'=&-3H-2\textbf{X}\cdot \bE \\
=&-\frac{1}{2}(3\textbf{X}'^2+4\textbf{X}\cdot\textbf{X}'')+6
\eea
This is the same as the statement that $f'=R+6$.  We note that $f$ is constant for any metric with Ricci scalar equal to $-6$, and is particular constant on a solution to the equations of motion.

Finally, we compute $\dot g_{\t\t}$, $\dot g_{\t\phi}$, $\dot g_{\phi\phi}$.  The resulting equations can be written as 
\bea\label{rc}
\dot \bX =&  \bY\times \bX +{1\over 2} f \bX' +2H {\bf X}+ X^2 \bE\\
=&\bY\times \bX +{1\over 2} f \bX' + 2\bX\left({1\over 2} \textbf{X}'^2 -{1\over \mu} \textbf{X}\cdot(\textbf{X}'\times\textbf{X}'')-2\right)\\
&+\bX^2 \left(\bX''+{1\over 2\mu}(2 \bX\times \bX'''+3 \bX'\times\bX'')\right)
\eea
Equations (\ref{rc}) and \rref{feq} are the differential equations that generate our flow.  They may be thought of as a set of modified heat equations for the vector field $\bX$.  We note that the right hand side of (\ref{rc}) contains parity odd terms which come from the Cotton tensor.

Before proceeding let us make a few comments.
We first note that the term involving $f$ in \rref{rc} describes the diffeomorphism $\sigma \to \sigma + const.$ generated by the $f \p_\sigma$ term in $v^\mu\p_\mu$.
The term involving $\bY\times \bX$ in \rref{rc} describes an infinitesimal $SO(2,1)$ rotation of $\bX$ around the $\bY$ axis; this rotation is induced by the components of the diffeomorphism $v^\mu\p_\mu$ in the $\t$ and $\phi$ direction.
As this diffeomorphism does not alter the geometry, one could simply set $\bY=0$.  However, we will find it useful to include this term in the next section when we study explicit flows.


\subsection{AdS and Warped AdS solutions}

Before proceeding it is useful to describe the known AdS and warped AdS solutions in the notation of the previous section. 

Both the AdS and warped AdS solutions are contained in the ansatz
\be
\bX = \bX_0 + \sigma \bX_1 + \sigma^2 \bX_2
\ee
The equation of motion
\be
\bE = 2 \bX_2 + {3\over \mu} \bX_1 \times \bX_2 = 0
\ee
implies that $\bX_2^2 = 0 = \bX_1\cdot \bX_2$.  We can then write the Hamiltonian constraint as
\be
H = {1\over 2} \bX_1^2 +{4\over 3} \bX_0 \cdot \bX_2-2= 0
\ee

There are two possible ways of solving the first equation.  The first is by setting $\bX_2=0$, in which case the equations imply that
\be
\bX_2=0~~~~\bX_1^2 = 4
\ee
This is AdS.  For example, setting $\bX_0=0$ gives AdS in Poincare' coordinates.
\be
ds^2 = {d\sigma^2\over 4 \sigma^2}-4\sigma d\tau d\phi
\ee
The second class of solutions have $\bX_2\ne0$.  In this case we have 
\be
\bX_2^2 = 0,~~~~\bX_1^2=\frac{4 \mu^2}{9},~~~~\bX_1\cdot\bX_2=0,~~~~\bX_0\cdot\bX_2={3\over 2}(1-\frac{\mu^2}{9})
\ee
These solutions describe warped AdS.  For example, the solution
\be
\bX = \left(\pm \sqrt{9-\mu^2\over 12}(\sigma^2-1), \pm \sqrt{9-\mu^2\over 12}(\sigma^2+1),{-2 \mu \over 3}\sigma \right)
\ee
describes warped AdS in planar-like coordinates.  The two signs describe spacelike and timelike warped AdS, respectively.

\section{Flows Connecting AdS and Warped AdS}

As described above, the massive graviton mode will cause the AdS solutions of TMG to decay.  The final state of this decay process will depend on the initial profile for the massive graviton; different perturbations of the AdS solution may lead to different end-states.  In general, we expect that there are two possible outcomes for this decay process.  The first is that the metric becomes singular, and the second is that the metric flows to warped AdS.   In this section we will show that both of these possibilities are realized, and that certain initial perturbations exist which take AdS to warped AdS, while others cause the space-time to decay completely.  We will do so by constructing families of metrics where the Ricci-Cotton flow is quite tractable.

\subsection{Minisuperspace Models}

In the previous section we simplified the Ricci-Cotton flow equations substantially by considering stationary, axisymmetric solutions.  Nevertheless, the equations remain quite complicated, as they describe a flow in the infinite dimensional space of functions $\bX(\sigma)$. 
We begin by asking under what circumstances it is possible to restrict the Ricci-Cotton flow to a finite dimensional space of metrics.  In general this is a difficult task; if we choose an arbitrary family of metrics, there is no guarantee that Ricci-Cotton flow will keep us inside this family.  

We will work in the sector of metrics with $U(1)\times U(1)$ symmetry and use the notation of the previous section.  We start by considering a family of metrics $\bX(\sigma, \lambda_i)$ parameterized by a finite number of variables $\l_i$ which are allowed to depend on the flow parameter $t$.  We will demand that Ricci-Cotton flow keeps us inside this finite dimensional space of metrics.  This means that the $t$-dependence of the metric can be completely incorporated into the $t$-dependence of $\l_i(t)$.  So
\be\label{minireq}
\dot \bX = \sum_i {d \bX \over d \l_i} \dot \l_i
\ee 
This equation is highly nontrivial, since it means that the vector $\dot \bX$ lies in the plane spanned by the ${d\bX \over d \l_i}$ for all values of $\sigma$.  Having found such a family, equating the two sides of \rref{minireq} yields a set of first order differential equations for the parameters $\lambda_i$.

In fact, by inspecting the forms of the AdS and warped AdS solutions in section 3.3 it is not difficult to find two-parameter families of metrics satisfying the condition (\ref{minireq}).  For these families, Ricci-Cotton flow reduces to a two dimensional flow.  We found several such choices which we present below; we will consider the cases $\mu>3$ and $\mu<3$ separately.

\subsection{Case $\mu>3$}

We consider the following family of metrics
\be\label{ansatz1}
\bX=\left(\l_1(\sigma^2+1),\l_1(\sigma^2-1), - 2 \l_2\sigma \right)
\ee
where $\lambda_i=(\lambda_1, \lambda_2)$ are our two parameters.  It is straightforward to check that condition (\ref{minireq}) is satisfied.  Plugging into the flow equation \rref{feq} we find that
\be
f=f_0+ 2\sigma(3+4 \l_1^2-3\l_2^2)
\ee
Computing $\dot \bX$ from \rref{rc} we find that 
\rref{minireq} is satisfied only if we choose
\be
\bY =\left(0,0,-3 -\lambda_2^2+ {12 \l_2^3\over \mu} - {12 \l_1^2\l_2\over \mu}  \right),~~~~~f_0=0
\ee
This choice of $t$-dependent diffeomorphism is necessary to enforce the gauge choice \rref{ansatz1} for all $t$.  With this value of $\bf Y$ the ${\dot \bX}$ equation implies that
\bea\label{fl1}
\dot \l_1 &= -\l_1(1-5 \l_2^2)-{4\over \mu}\l_1\l_2(\l_1^2+3\l_2^2) \\
\dot\l_2 &= -\l_2(1-4\l_1^2 -\l_2^2) - {16\over \mu}\l_1^2\l_2^2
\eea 
This is the first order flow on the two dimensional space of metrics \rref{ansatz}.
This flow can not be solved in terms of simple closed form expressions, but nevertheless all of the important features of the flow can be understood from \rref{fl1}.

The advantage of this two-parameter subspace is that it contains AdS and warped AdS as fixed points.  Comparing with section 3.3 we see that AdS and warped AdS are given by
\bea\label{sl1}
(\l_1,\l_2)_{AdS} &= (0,1) \\
(\l_1,\l_2)_{SAdS} &= \left(\sqrt\frac{(\mu^2-9)}{12},{ \mu\over 3}\right)\\
(\l_1,\l_2)_{TAdS} &= \left(-\sqrt\frac{(\mu^2-9)}{12},{ \mu\over 3}\right)
\eea
The latter two denote Spacelike and Timelike warped AdS, respectively.  
It is easy to check that these solutions are fixed points of the flow \rref{fl1}. By linearizing  the flow \rref{fl1} around  these solutions one can check that the solutions \rref{sl1} are all unstable against a general perturbation in the $\lambda_1-\lambda_2$ plane.  In particular, the AdS solution is completely unstable as the Hessian matrix is negative definite, while the warped AdS vacua are saddle points with indefinite Hessians. This does not come as a surprise; the family of metrics  \rref{ansatz1} have not been chosen to preserve any particular boundary conditions and the solutions are not necessarily expected to be stable.
We note also that this family of metrics includes singular configurations, such as $(\l_1,\l_2)=(0,0)$, which are fixed points of the flow as well.

\begin{figure}[htbp]\label{f1}
\centering
\includegraphics[totalheight=0.25\textheight,width=0.5\textwidth]{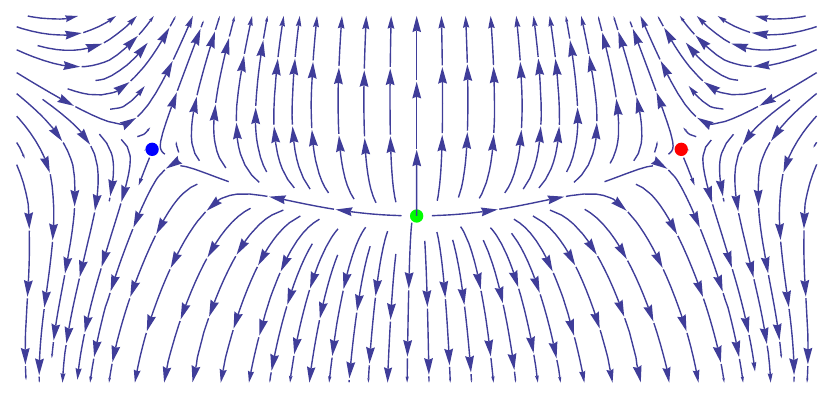}
\caption{Typical flow for the case $\m>3$.  The green dot (in the center of the diagram) is the AdS solution, the red dot (on the right side of the diagram) is spacelike warped AdS and the blue dot (on the left side of the diagram) is timelike warped AdS.}
\end{figure}

The important point, however, is that the flow \rref{fl1} connects the AdS and warped AdS solutions.  This can be seen in Figure 1, where the flow is plotted.  We see that there is a particular perturbation of the AdS solution which will cause it to decay into warped AdS.  This perturbation must be finely tuned, or else the flow will miss the warped AdS fixed point and flow to a singular solution; this is a consequence of the fact that the warped AdS vacua are saddle points in this parameterization, rather than stable minima.\footnote{This is a common occurrence in the study of tachyon condensation.  For example, in the study of open string tachyon condensation a generic perturbation of a brane-anti brane configuration will cause the solution to decay completely, but certain finely tuned initial perturbations will cause the system to decay to an array of lower dimensional branes (see \cite{Sen:2004nf} for a review).}  Finally, we note that although there is no closed form analytic expression for the curve in the $\lambda_1-\lambda_2$ plane which describes this flow, the existence of such a flow is guaranteed by the smoothness of the equations \rref{fl1}.

We note that the results presented above depend on the choice of ansatz \rref{ansatz1}.  Many other choices are possible.  We have found others, which we omit for brevity, as the results are essentially the same as those presented above. 

We emphasize that the flows considered above do not respect the standard asymptotically AdS boundary conditions, nor do they respect the warped AdS boundary conditions proposed in \cite{Anninos:2009zi}, for which spacelike stretched AdS is argued to be stable.  This is, perhaps, the main drawback of the Ricci-Cotton flow technique; a family of metrics which connects AdS to warped AdS will, almost by definition, violate AdS boundary conditions.  It would be interesting to find a flow which violates AdS boundary conditions, but lies inside the family of warped AdS metrics proposed by \cite{Anninos:2009zi}.  We leave this as an interesting open question.

\subsection{Case $\mu<3$}

In this case the ansatz for the metric must be modified.  We take
\be
\bX=\left(\l_1(\sigma^2-1),\l_1(\sigma^2+1), 2 \l_2\sigma \right)
\ee
In this case the ansatz is preserved if 
\be
f= 2\sigma(3-4 \l_1^2-3\l_2^2),~~~~\bY =\left(0,0,-3 -\lambda_2^2-{12 \l_2^3\over \mu} - {12 \l_1^2\l_2\over \mu}\right)
\ee 
The flow equations are
\bea
\dot \l_1 &= -\l_1(1-5 \l_2^2)-{4\over \mu}\l_1\l_2(\l_1^2-3\l_2^2) \\
\dot\l_2 &= -\l_2(1+4\l_1^2 -\l_2^2) - {16\over \mu}\l_1^2\l_2^2
\eea 
Again, AdS, Spacelike warped AdS and Timelike warped AdS appear as fixed points: 
\bea
(\l_1,\l_2)_{AdS} &= (0,\pm1) \\
(\l_1,\l_2)_{SAdS} &= \left(\sqrt\frac{(9-\mu^2)}{12},-{ \mu\over 3}\right)\\
(\l_1,\l_2)_{TAdS} &= \left(-\sqrt\frac{(9-\mu^2)}{12},-{ \mu\over 3}\right)
\eea
In this case the warped AdS vacua are completely unstable, while the AdS solution is a saddle point.

\begin{figure}[htbp]\label{f2}
\begin{center}
\includegraphics[totalheight=0.25\textheight,width=0.5\textwidth]{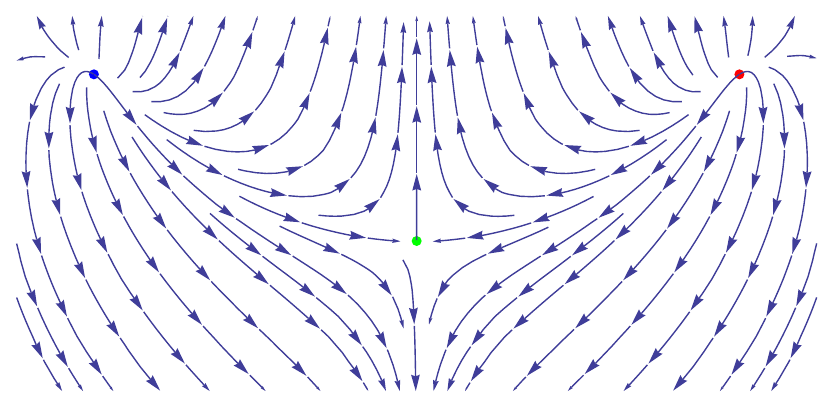}
\caption{Typical flow for the case $\m<3$.  The green dot (center) is the \ads solution, the red dot (on the right) Spacelike warped AdS and the blue dot (on the left) Timelike warped AdS.}
\end{center}
\end{figure}
The flow is plotted in Figure 2. 
We see that there is a perturbation of the warped AdS solutions which will cause them to decay into AdS.  This perturbation must be finely tuned, or else the flow will miss the AdS completely, just as in the previous case.  Again, although there is no closed form analytic expression which describes this flow, the existence of a solution is guaranteed.  

Unlike the $\mu>3$ case, however, the physical interpretation of this result is less clear, as there is no conjectured stable ground state of TMG for $\mu<3$.

\

In conclusion, we have demonstrated that Ricci-Cotton flows exist which explicitly connect AdS and warped AdS.  For values of $\mu>3$ -- which are precisely the cases where warped AdS possesses a conjectured CFT dual -- AdS will decay to warped AdS under certain perturbations.  This indicates that AdS, although it is unstable, may exist as an unstable "sphaleron" state in the CFT dual to warped AdS.

 \section*{Acknowledgements}
This work is
supported by the National Science and Engineering Research Council
of Canada.  We wish to thank A. Castro, J. Gegenberg, T. Hartman,  M. Headrick, W. Song, A. Strominger and A. Wissanji
for useful conversations.

\end{document}